\documentstyle[12pt,aaspp4,epsf]{article}
\newcommand{\beq}{\begin{equation}}
\newcommand{\beqa}{\begin{eqnarray}}
\newcommand{\eeq}{\end{equation}}
\newcommand{\eeqa}{\end{eqnarray}}

\newcommand{\siml}{\lesssim}

\newcommand{\bm}[1]{\mbox{{\boldmath $#1$}}} 
\begin{document}
%
%{KUNS-1711,YITP-}
\title{
 Peak Luminosity-Spectral Lag Relation 
 Caused by the Viewing Angle of the Collimated Gamma-Ray Bursts
}
\author{
Kunihito Ioka$^{1}$
and Takashi Nakamura$^{2}$
}
\affil{$^{1}$ Department of Physics, Kyoto University, Kyoto 606-8502,
Japan}
\affil{
$^{2}$Yukawa Institute for Theoretical Physics, Kyoto University, 
Kyoto 606-8502, Japan}
\centerline{Mar~~2~~2001}
\authoremail{iokakuni@tap.scphys.kyoto-u.ac.jp, 
takashi@yukawa.kyoto-u.ac.jp}

\begin{abstract}
We compute the kinematical dependence of 
the peak luminosity, the pulse width and the spectral lag of the peak
luminosity on the viewing angle $\theta_v$ of a jet.
For appropriate model parameters we obtain the 
peak luminosity-spectral lag relation similar to the observed one
including GRB980425.
A bright (dim) peak with short (long) spectral lag corresponds to a jet
with small (large) viewing angle.
This suggests that the viewing angle of the jet
might cause various relations in GRBs
such as the peak luminosity-variability relation
and the luminosity-width relation.
Our model also suggests that X-ray rich GRBs 
(or X-ray flushes or Fast X-ray transients)
are typical GRBs observed from large $\theta_v$
with large spectral lag and low variability.
\end{abstract}

\keywords{gamma rays: bursts --- gamma rays: theory}

\section{INTRODUCTION}
Afterglows of the GRBs are believed to be produced by the
external shocks while the GRB itself is by the internal shocks.
There is a growing evidence that some of afterglows, such as GRB000926,
GRB000301C, GRB990510 and GRB990123,
are collimated with an opening half-angle of $\Delta \theta^{(a)} \sim 0.1$.
However, the collimation half-angle $\Delta \theta$
of the internal shocks remains unknown
since we can observe only the angular region inside 
the relativistic beaming half-angle $\sim \gamma^{-1}$ where
$\gamma$ is the Lorentz factor of the flow and is larger than 
$\sim 100$ to avoid the compactness problem (e.g., Piran 1999).
Since the angular size of a causally connected region 
is also $\sim \gamma^{-1}$,
the minimum possible $\Delta \theta$ is $\sim \gamma^{-1}
\siml 10^{-2} \ll \Delta \theta^{(a)}$, that is,  
$\Delta \theta$ can be much smaller than $\Delta \theta^{(a)}$.

We here assume that the internal shocks consist
of such sub-jets with the collimation half-angle of
$\Delta \theta \ll \Delta \theta^{(a)}$.
If we happen to observe a bright sub-jet,
the isotropic luminosity becomes very large so that
the sub-jets model can interpret 
(1) no correlation between the gamma-ray fluence and the redshift, 
(2) the weak correlation between the gamma-ray fluence and the afterglow flux
as well as (3) the extremely energetic bursts $\sim 10^{54}$ ergs
(Kumar \& Piran 2000; but see also Freedman \& Waxman 1999).
The X-ray pre/post-cursor (Murakami et al. 1991; Sazonov et al. 1998;
in't Zand et al. 1999) can also be explained in the sub-jets model, 
if some of off-axis sub-jets are ejected earlier (for precursor) 
or later (for postcursor) than the main sub-jets (Nakamura 2000).
X-ray rich GRBs (Strohmayer et al. 1998;
Frontera et al. 2000; Heise et al. 2001; Kippen et al. 2001)
are also expected in the sub-jets model 
if none of sub-jets point to our line-of-sight (Nakamura 2000).
The ``sub-jet'' is sometimes termed ``cannonball'' 
(Dar \& De R${\acute {\rm u}}$jula 2000a, b; Plaga 2000),
``mini-jet'' (Shaviv \& Dar 1995) and so on.

On the other hand, there are some possible relations 
between the peak luminosity, the
spectral lag, the pulse width and the variability of GRBs.
{}From six GRBs with known redshifts, Norris, Marani \& Bonnell (2000) found 
that the isotropic peak luminosity is inversely proportional 
to the spectral lag which is defined as the time lag of the peak 
luminosity between different energy bands.
Fenimore \& Ramirez-Ruiz (2000)
found that the luminosities of seven bursts with known redshifts are
correlated with the variabilities of the bursts 
(see also Reichart et al. 2000).
If both relations are correct, there should be a relation between
the variability and the spectral lag.
Recently Schaefer, Deng \& Band (2001) showed
that this is the case by plotting the variability and 
the spectral lag of 112 GRBs.
Therefore at present there is a strong support that
the luminosity-spectral lag as well as
the luminosity-variability relation are correct. 
 
There are several attempts to explain the luminosity-spectral lag and
the luminosity-variability relations.
Salmonson (2000, 2001) proposed that it is
due to the difference of the observed cooling or
deceleration time which depends on
the variation in line-of-sight velocity among bursts.
Schaefer (2001) interpreted the luminosity-spectral lag 
and luminosity-variability relation
as a consequence of the conservation of energy when radiative cooling
dominates and the dependence of the luminosity and variability
on the Lorentz factor of the ejecta, respectively.
However Wu \& Fenimore (2000) pointed out that the synchrotron cooling time
scale is much shorter than the lag time scale.
Plaga (2000) argued that the luminosity-variability relation
might be obtained in the framework of the cannonball model
for the given power low power density spectrum of the GRBs.

In this Letter, under the assumption of short cooling time scale, 
we will show that depending on the viewing angle of a single jet 
to our line-of-sight, the luminosity-spectral lag and 
the luminosity-variability relation may arise kinematically.
In the sub-jets model, we at present do not know the number of
sub-jets in $\Delta \theta^{(a)}$. If it is small, that is, the
filling factor of the sub-jets is small, we should study effects of
many independent sub-jets based on the present results of a single
jet. If the number of sub-jets is large, we may apply the present results
regarding $\Delta \theta \sim \Delta \theta^{(a)}$.
In any case it is important to study the viewing angle dependence of
various physical quantities, which is the main purpose of this Letter
irrespective of the details of the sub-jets model.
%In \S \ref{sec:emission} we will formulate the instantaneous emission 
%from an expanding infinitesimally thin shell
%with a finite opening angle, i.e., from a jet.
%In \S \ref{sec:profile}, we will investigate the properties of 
%a single pulse produced by a jet.
%In \S \ref{sec:lumilag} and \S \ref{sec:lumivari},
%we will explore the luminosity-spectral lag and 
%the luminosity-variability relation, respectively.
%\S \ref{sec:discuss} is devoted to discussions.

\section{INSTANTANEOUS EMISSION FROM AN EXPANDING THIN JET}\label{sec:emission}
The internal shock occurs in the relativistic wind
when the fast moving flow catches up the slow one.
The wind can be modeled by a succession of relativistic shells
(Kobayashi, Piran \& Sari 1997; Daigne \& Mochkovitch 1998).
A collision of two shells produces a single pulse light curve, whose
superposition makes a whole light curve. 
Here we consider a single pulse.
There are three time scales that determine the temporal pulse structure:
the hydrodynamic time scale, the cooling time scale,
and the angular spreading time scale 
(Kobayashi, Piran \& Sari 1997; Katz 1997; Fenimore, Madras \& Sergei 1996).
The cooling time scale is usually much shorter than 
the hydrodynamic time scale in the internal shocks
(Sari, Narayan \& Piran 1996),
so that we may assume a delta function time dependence of the emissivity
in the comoving frame.
Let $l$ and $L$ be the width of the shell and  the separation of the shells.
Then the hydrodynamic time scale $\sim l/c$ and the angular spreading
time scale $\sim L/c$ determine 
the rise and the decay time of a single pulse, respectively.
Since most observed pulses rise more quickly than they decay
(Norris et al. 1996), for simplicity we may adopt an infinitesimally thin shell
approximation as $l \ll L$.
%\footnote{
%If $l$ is not much smaller than $L$,
%we can use the results of the infinitesimally thin shell model 
%for further study since we can regard that the shell is made of a
%number of infinitesimally thin shells.
%Although this model neglects pressure waves,
%it can still capture the basic feature of the real process
%(Daigne \& Mochkovitch 1998, 2000).}

Let us consider an emitting thin shell
that is confined to a cone of an opening
half-angle $\Delta \theta$ and moves radially outward 
with the Lorentz factor of $\gamma=1/\sqrt{1-\beta^2}$.
A general formula to calculate the observed flux  
from an optically thin material
is derived by Granot, Piran \& Sari (1999) and Woods \& Loeb (1999).
Here we adopt their formulations and notations.
Let us use a spherical coordinate system ${\bm r}=(r, \theta, \phi)$
in the lab-frame, where the $\theta=0$ axis points toward the detector
and the central engine is located at the origin.
Let also the detector be at a distance $D$ from the source
and $\alpha:=r\sin\theta/D=r\sqrt{1-\mu^2}/D$ be the angle that a given 
ray makes
with the $\theta=0$ axis.
Then the observed flux at the observed time $T$, 
measured in erg s$^{-1}$ cm$^{-2}$ Hz$^{-1}$,
is given by
$F_{\nu}(T)={{\nu D}\over{\gamma \beta}}
\int^{2\pi}_{0} d\phi \int^{\alpha_m}_0 \alpha^2 d\alpha
\int^{\nu \gamma(1+\beta)}_{\nu \gamma (1-\beta)}
{{d\nu'}\over{{\nu'}^2}}
{{j'_{\nu'} \left(\Omega_d', {\bm r}, t\right)}
\over{(1-\mu^2)^{3/2}}}$
and
$\mu=(1-\nu'/\gamma \nu)/\beta$,
where $\Omega_d'$, $t=T+(r\mu/c)$, $\alpha_m$,  and $j'_{\nu'}$
are the direction towards the detector measured in the comoving frame,
the lab-frame time,
the maximum value of $\alpha$,
and the comoving frame emissivity 
in units of ergs s$^{-1}$ cm$^{-3}$ Hz$^{-1}$ sr$^{-1}$.
Note here that a prime means the physical quantities in the comoving frame.

If the emission is isotropic in the comoving frame,
the emissivity has a functional form of
$j'_{\nu'}(\Omega_d', {\bm r}, t)=A_0 f(\nu') \delta(t-t_0) 
\delta(r-r_0) H(\Delta \theta-|\theta-\theta_v|)
H\left[\cos \phi-\left({{\cos\Delta \theta-\cos\theta_v \cos\theta}\over
{\sin \theta_v \sin\theta}}\right)\right]$,
where  $f(\nu')$ represents the spectral shape and
$\theta_v$ is the angle that the axis of the emission cone makes
with the $\theta=0$ axis. 
The delta functions describe
an instantaneous emission at $t=t_0$ and $r=r_0$,
and $H(x)$ is the Heaviside step function which
describes that the emission is inside the cone.
%Using the property of the delta function,
%the relation $\alpha=r\sqrt{1-\mu^2}/D$ and equation (\ref{eq:mu}),
%we can change the delta functions of $t$ and $r$
%into those of $\nu'$ and $\alpha$, respectively.
%Then, equations (\ref{eq:fnugeneral}), (\ref{eq:mu}) and 
%(\ref{eq:jnuthin}) yield
Then, the flux of a single pulse is given by
\beqa
F_{\nu}(T)
={{2 c A_0 r_0 \gamma^2}\over{D^2}}
{{\Delta \phi(T) f\left[\nu\gamma (1-\beta\cos\theta(T))\right]
}\over{\left[\gamma^2 (1-\beta\cos\theta(T))\right]^2}},
\label{eq:jetthin}
\eeqa
where
$1-\beta\cos\theta(T)=({c\beta}/{r_0})(T-T_0)$
and $T_0=t_0-r_0/c\beta$.
For $\Delta \theta>\theta_v ~{\rm and}~ 0<\theta(T)\le \Delta
\theta - \theta_v$, $\Delta \phi(T)=\pi$,
otherwise 
$\Delta \phi(T)=
\cos^{-1}\left[
{{\cos \Delta \theta - \cos \theta(T) \cos \theta_v}\over
{\sin \theta_v \sin \theta(T)}}\right]$.
For $\theta_v < \Delta \theta $, $\theta(T)$ varies from 0 to 
$\theta_v+\Delta \theta$ while from $\theta_v-\Delta \theta$ to
$\theta_v+\Delta\theta$ for $\theta_v > \Delta \theta $. 
In the latter case, $\Delta \phi(T)=0$ for $\theta(T)=\theta_v-\Delta \theta$.

The observed spectrum of GRBs is well approximated by 
the Band spectrum (Band et al. 1993). In order to 
have a spectral shape similar to the Band spectrum,
we adopt the following form of the spectrum in the comoving frame,
\beqa
f(\nu')=\left({{\nu'}\over{\nu'_0}}\right)^{1+\alpha_B}
\left[1+\left({{\nu'}\over{\nu'_0}}\right)^{s}\right]^{(\beta_B-\alpha_B)/s},
\label{eq:spectrum}
\eeqa
where $\alpha_B$ ($\beta_B$) is the low (high) energy power law index,
and $s$ describes the smoothness of the transition between
the high and low energy.  
$\alpha_B\sim -1$ and $\beta_B\sim -2.5$ are typical values
(Preece et al. 2000).
Equations (\ref{eq:jetthin})
%(\ref{eq:theta}), (\ref{eq:phi}) 
and (\ref{eq:spectrum})
are the basic equations to calculate the flux of a single pulse,
which depends on nine parameters
for $\theta_v\ll 1$ and $\Delta \theta \ll 1$:
$\gamma \nu_0'$, $\gamma \theta_v$, $\gamma \Delta \theta$,
$r_0/c \beta \gamma^2$, $T_0=t_0-r_0/c\beta$, $\alpha_B$,
$\beta_B$, $s$ and $2 c A_0 r_0 \gamma^2/D^2$.

\section{PULSE PROFILE AND SPECTRUM}\label{sec:profile}
In order to study the viewing angle dependence of a pulse from a
single jet,
we first fix other parameters:
$\gamma \Delta \theta=1$, $\alpha_B=-1$, $\beta_B=-2.5$ and $s=1$.
The lab-frame frequency, the observed time and the flux per frequency
are measured in units of $\gamma \nu_0'$,
$r_0/c \beta \gamma^2$ and $2 c A_0 r_0 \gamma^2/D^2$, respectively.
The total fluence is given by
$S_{\nu}=\int_{T_{start}}^{T_{end}} F_{\nu}(T) dT$,
where
$T_{start}=T_0+({r_0}/{c\beta})
(1-\beta\cos(\max[0,\theta_v-\Delta\theta]))$
and
$T_{end}=T_0+({r_0}/{c\beta})(1-\beta\cos(\theta_v+\Delta\theta))$.
In Figure 1, we plot  ${\nu} S_{\nu}$
as a function of the observed frequency $\nu$
by varying the viewing angle $\gamma \theta_v$.
As $\gamma \theta_v$ increases, both the maximum frequency
and ${\nu} S_{\nu}$ decrease. 
Here the maximum frequency $\nu_{max}$ means the
frequency at which most of the radiation energy is emitted.
In Figure 1 we also show the analytically estimated dependence,
$\nu_{max}\propto \delta^{-1}$ and $(\nu S_{\nu})_{max}\propto \delta^{-3}$,
by triangles, where $\delta=\gamma(1-\beta\cos\theta_v)\simeq
(1+\theta_v^2\gamma^2)/2\gamma$ 
is the Doppler factor 
%(see Appendix \ref{sec:analytical} for
(see Ioka \& Nakamura 2001 for
analytical estimates.).
%We see the good agreement between triangles and computed maxima.
Figure 1 shows that if we observe the jet from
$\theta_v \sim {\rm several} \times \gamma^{-1}$ the maximum 
frequency $\nu_{max}$ is in the X-ray
band while for $ \theta_v \sim \gamma^{-1}$ it is in the gamma-ray
region so that depending on the viewing angle, GRBs may be
observed as X-ray rich GRBs in our model (Nakamura 2000).
Note that if the emitting sub-jets distribute
sparsely within the opening angle of the afterglow $\Delta \theta^{(a)}$
or if we are seeing the edge of the afterglow cone, the afterglow
can occur as usual.

Several pulse profiles are shown in the upper panel of Figure 2,
where the observed frequency is taken as 
$\nu=200{\rm keV}(\gamma \nu_0'/10^3{\rm keV})$.
The pulse profiles are similar to the FRED (Fast Rise Exponential Decay)
shape observed in many bursts.
Note here that the vertical axis is a log scale so that
the straight line means the exponential decay.
In the lower panels of Figure 2 we show 
the pulse width $W_{FWHM}$ at half maximum and 
the product of the peak flux and the width
$\nu F_{\nu}^{peak} W_{FWHM}$ as a function
of $(1+\gamma^2\theta_v^2) \propto \delta$ 
with the analytically estimated dependence on $(1+\gamma^2\theta_v^2)$
%(see Appendix \ref{sec:analytical}). 
(see Ioka \& Nakamura 2001). 
%We see the good agreement between the computed
%results and analytical estimates in each region of 
%$(1+\gamma^2\theta_v^2)$.
%Therefore 
Figure 2 shows that
the luminosity-width relation is given by
$\nu F_{\nu}^{peak}\propto W_{FHWM}^{\kappa}$
where $\kappa=-2+\alpha_B\sim -3$ for $\theta_v \sim \Delta \theta$
and $\gamma \theta_v \siml [10(\gamma\nu_0'/10^3 {\rm keV})
(\nu_{obs}/200 {\rm keV})^{-1}-1]^{1/2}$.
This is consistent with the observation
that the narrower pulse tends to be brighter in each burst
following $\nu F_{\nu}^{peak}\propto W_{FWHM}^{-2.8}$
(Ramirez-Ruiz \& Fenimore 2000).
Our calculations predict that 
the exponent $\kappa$ ranges from $\kappa=-2+\alpha_B\sim -3$
to $\kappa=-3+2\beta_B \sim -8$
for dimmer bursts including X-ray rich GRBs which have the larger viewing
angle in our model.

\section{PEAK LUMINOSITY-SPECTRAL LAG/VARIABILITY
RELATION}\label{sec:lumilag}
In the upper panel of Figure 3, varying the viewing angle 
$\gamma \theta_v$ 
we show the isotropic peak luminosity 
at frequency $\nu_{\gamma}=200{\rm keV}(\gamma \nu_0'/10^3{\rm keV})$
as a function of the spectral lag in an arbitrary vertical scale 
of $2 c A_0 r_0 \gamma^2/D^2$.
For simplicity, we define the spectral lag $\Delta T$
as the difference of the peak time
between frequencies $\nu_{\gamma}=200{\rm keV}(\gamma \nu_0'/10^3{\rm keV})$
and $\nu_{X}=20{\rm keV}(\gamma \nu_0'/10^3{\rm keV})$,
although the spectral lag is determined by the cross-correlation
analysis in Norris, Marani \& Bonnell (2000).
Since there is no lag when 
the observer is inside the emission cone $\theta_v<\Delta \theta$
in our present definition,
all points with $\Delta T\le 10^{-4}(r_0/c \beta \gamma^2)$
are plotted at $\Delta T=10^{-4}(r_0/c \beta \gamma^2)$.
In the lower panel of Figure 3 the solid line shows 
the corresponding viewing angle $\gamma\theta_v$
as a function of $\Delta T$.
We adopt $r_0/c\beta \gamma^2=10$ s.
The peak luminosity first decreases with increasing spectral lag
and turns back for large spectral lag
since the different $\theta_v$ give the same lag.
The analytical estimates, $\nu F_{\nu}^{peak}\propto \Delta T^{-3/2}$
and $\gamma \theta_v \propto \Delta T^{1/4}$,
shown by dotted lines agree well with the solid lines 
for $1\siml \gamma \theta_v \siml 3$ 
%(see Appendix \ref{sec:analytical}).
(see Ioka \& Nakamura 2001).
The deviation from the power law at low luminosity
is due to the fact that the maximum frequency $\nu_{max}$ 
becomes lower than the observed gamma-ray frequency $\nu_{\gamma}$
(see Figure 1).
The observed luminosity-spectral lag relations for seven bursts
with known redshifts (Norris, Marani \& Bonnell 2000) are also plotted
for comparison. Surprisingly enough, a simple sub-jet model happens
to reproduce the observation quite well including GRB980425 which has 
the extremely dim luminosity and large spectral lag.
%Although the real GRBs may be produced by more complicated 
%multiple sub-jets than a single sub-jet in our model of
%this Letter, this agreement suggests that the
%viewing angle of the sub-jets may be responsible for 
%the peak luminosity-spectral lag relation.

%\section{PEAK LUMINOSITY-VARIABILITY RELATION}\label{sec:lumivari}
The sub-jets model might be compatible with
the luminosity-variability relation  
that more complex bursts tend to be brighter
(Fenimore \& Ramirez-Ruiz 2000; Reichart et al. 2000;
Schaefer, Deng \& Band 2001).
According to Plaga (2000), when the power density spectrum of 
the GRB time histories is $P\propto f^{d}\sim f^{-5/3}$ 
(Beloborodov, Stern \& Svensson 1998, 2000; Shaviv \& Dar 1995),
the variability relates to the pulse width as
$V\propto \delta^{d+1} \propto W_{FHWM}^{d+1}$.
%{}From Figure 2 and Appendix \ref{sec:analytical},
{}From Figure 2 and \S \ref{sec:profile},
the peak flux at a given frequency depends on the pulse width
as $\nu F_{\nu}^{peak}\propto W_{FHWM}^{\kappa}$.
%where $\kappa=-2+\alpha_B\sim -3$ for the brightest bursts
%and $\kappa=-3+2\beta_B\sim -8$ for the dimmest bursts.
Therefore
the luminosity-variability relation is given by
$\nu L_{\nu}^{peak} \propto V^{\kappa/(d+1)} \sim V^{4.5} (V^{12})$
for the brightest (dimmest) bursts.
In the upper panel of Figure 4, we show
 $\nu L_{\nu}^{peak} \propto V^{\kappa/(d+1)} $
by the solid line , using $\nu L_{\nu}^{peak}$ and $W_{FHWM}$ 
in Figure 2 and \S \ref{sec:profile} with  appropriate off-sets.
%The corresponding viewing angle $\gamma \theta_v$ is shown 
%in the lower panel.
The observed luminosity-variability relations for eight bursts
with known redshifts (Fenimore \& Ramirez-Ruiz 2000) are also plotted
for comparison. 
Surprisingly again, a simple sub-jet model happens
to reproduce the observation quite well including GRB980425
which has the extremely low variability.
Although the power density spectrum of the GRBs has been assumed,
this agreement together with the results of the luminosity-spectral lag 
relation suggests that a single sub-jet model may capture the main features
of the GRBs.

\section{DISCUSSIONS}\label{sec:discuss}
GRB980425 and its apparent association with SN1998bw
first provided a possible connection between GRBs and SNe.
Since GRB980425 was an unusual GRB,
it may be a new type of GRB.
However, we have shown that the unusually low luminosity, large spectral lag
and low variability of GRB980425
can be interpreted by the off-axis sub-jet emission with
the viewing angle of $\theta_v \sim 10/\gamma \sim 6^{\circ}$.
The off-axis sub-jet emission may also explain why
GRB980425 was not seen in BATSE's highest energy band ($>300$keV;
Norris, Marani \& Bonnell 2000),
since the maximum frequency $\nu_{max}$ 
becomes lower than the observed frequency from Figure 1.
Furthermore, the unusually slowly declining X-ray
afterglow of GRB980425 ($\propto T^{-0.2}$)
can also be explained by the off-axis emission from the collimated afterglow
with the viewing angle of $\theta_v^{(a)} \sim 30^{\circ}$
(Nakamura 1999).
Since the axis of the afterglow can differ from that of the sub-jet,
$\theta_v \sim 6^{\circ}$ is compatible with
$\theta_v^{(a)} \sim 30^{\circ}$ 
if the sub-jets on the edge of the afterglow were seen.
Moreover from the late time light curve of SN1998bw,
it is suggested that the explosion is jet-like and our line-of-sight
is off-axis from the jet (Maeda et al. 2000).
Therefore GRB980425 may be a typical GRB observed from the large
viewing angle so that our results have strengthened the association 
of GRBs with SNe.

%One may consider that off-axis emission in the internal shocks 
%leads the lack of the afterglow.
%However if the emitting sub-jets distribute
%sparsely within the opening angle of the afterglow $\Delta \theta^{(a)}$
%or if we are seeing the edge of the afterglow cone, the afterglow
%can occur as usual.
%The sparsely distributed sub-jets model is compatible
%with the recent detection of many dim X-ray rich GRBs 
%(Heise et al. 2001; Kippen et al. 2001), 
%since the maximum frequency $\nu_{max}$ of off-axis emission 
%can be in the X-ray range (Nakamura 2000; see Figure 1).
%This interpretation predicts that X-ray rich GRBs
%have large spectral lag with low variability.

The bright bursts, such as GRB990123, 
can be suspected of on-axis emission with no or little spectral lag 
in our model,
so that observation with fine time resolution is important
to measure very small lag.
In reality, a burst should have more or less off-axis emission 
so that the lag should exist in the cross-correlation analysis
(Norris, Marani \& Bonnell 2000; Wu \& Fenimore 2000).
In this Letter we have studied only the behavior of the single sub-jet under
the simple treatment of the spectral lag.
Nevertheless we have obtained theoretical results very similar to the
observations, which  suggests the viewing angle is one of the
most important factors to determine the various relations in GRBs.
In the next step we should consider the multiple sub-jets case
using the present results of the single sub-jet.\footnote{
When we almost completed this Letter, Frail et al. (2001)
suggested that GRBs with the narrower $\Delta \theta^{(a)}$ tend
to have larger isotropic gamma-ray total energy. 
In our model this suggests that the filling factor of the sub-jets within 
$\Delta \theta^{(a)}$ is another key parameter.}

\acknowledgments
We are grateful to P. M${\acute {\rm e}}$sz${\acute {\rm a}}$ros
for useful comments.
KI would like to thank H. Sato for continuous
encouragement and useful discussions.
This work was supported in part by
Grant-in-Aid for Scientific Research Fellowship
of the Japanese Ministry of Education,
Science, Sports and Culture, No.9627 (KI) and by
Grant-in-Aid of Scientific Research of the Ministry of Education,
Culture, and Sports, No.11640274 (TN) and 09NP0801 (TN).

\newpage 
\begin{figure}
    \epsfysize 17cm 
    \epsfbox{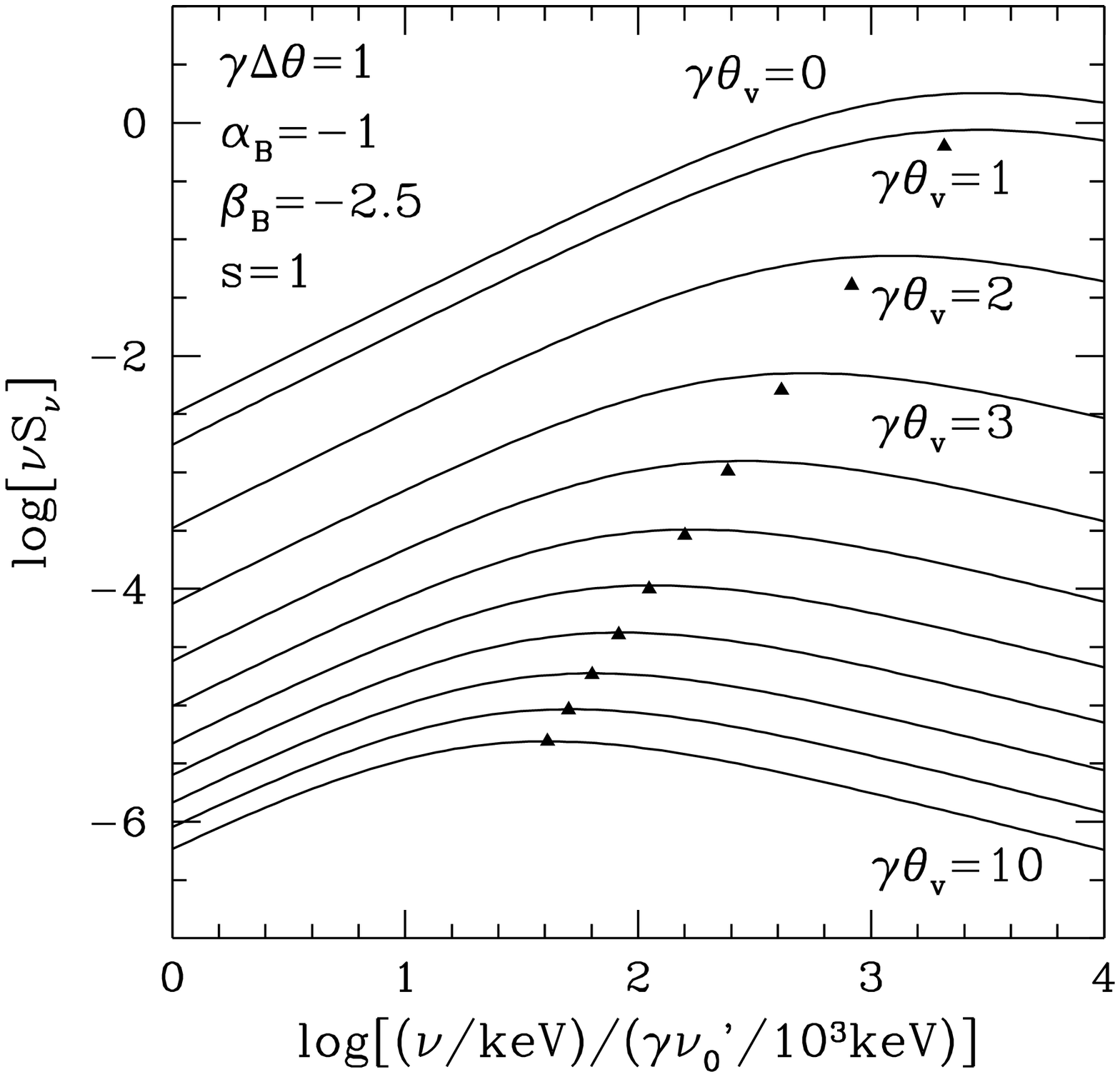}
\caption[fig1.ps]
%\figcaption
{The total fluence ${\nu} S_{\nu}$ in units of $2A_0 r_0^2 \gamma \nu_0'/
\beta D^2$ is shown 
as a function of the observed frequency
by varying the viewing angle $0\le \gamma \theta_v\le 10$, 
($\gamma \theta_v$: integer).
We adopt $\gamma \Delta \theta=1$,
$\alpha_B=-1$, $\beta_B=-2.5$ and $s=1$.
The triangles represent the analytically estimated dependence,
$\nu_{max} \propto \delta^{-1}$ and $(\nu S_{\nu})_{max} \propto \delta^{-3}$,
where $\delta=\gamma(1-\beta \cos \theta_v) \simeq 
(1+\theta_v^2 \gamma^2)/2\gamma$ is the Doppler factor
and the normalization is set by the peak of $\gamma \theta_v=10$.}
\end{figure}

\newpage 
\begin{figure}
    \epsfysize 17cm 
    \epsfbox{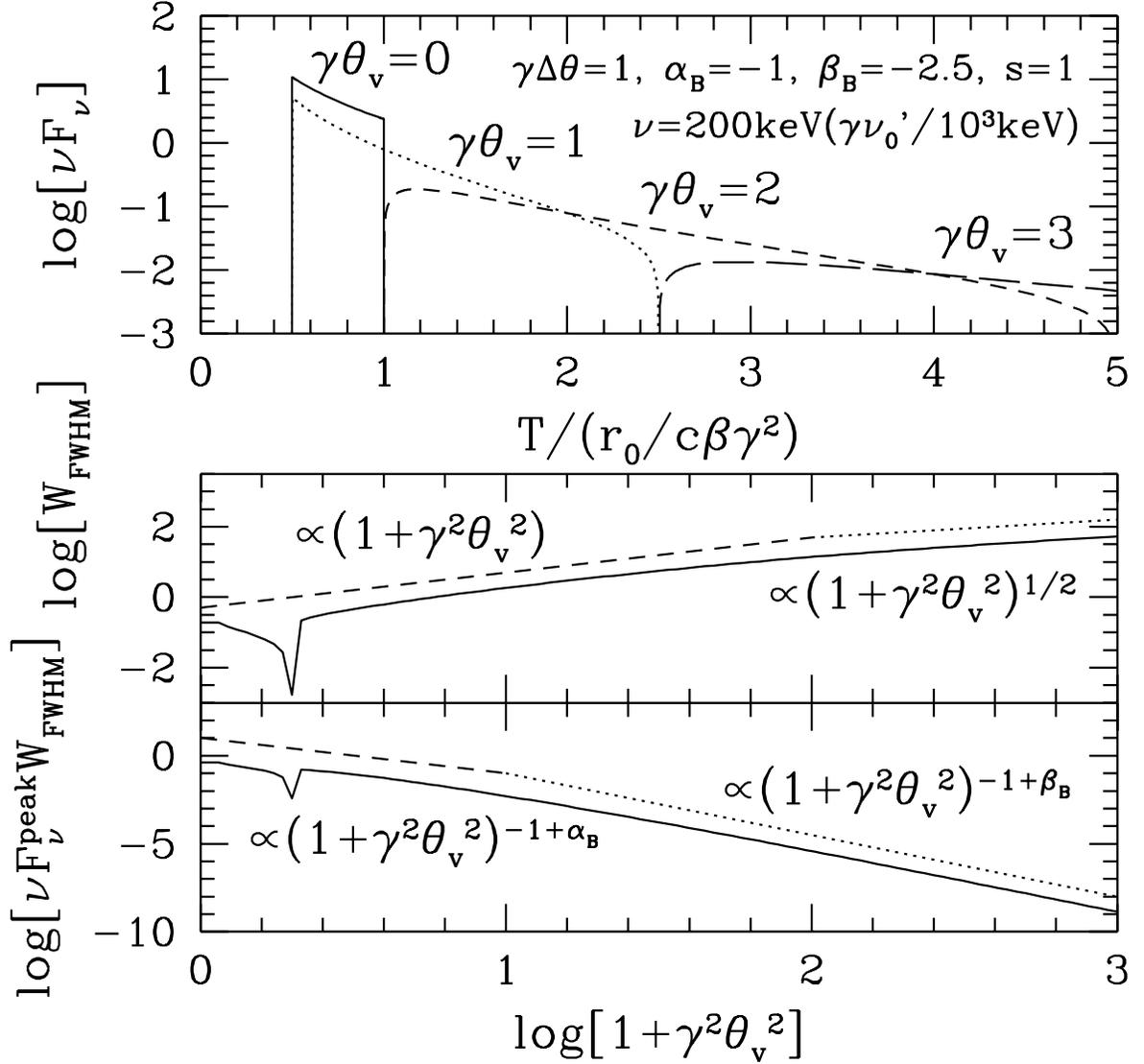}
\caption[fig2.ps]
%\figcaption
{({\it upper panel}): Pulse profiles are shown.
The pulse begins at 
$T_{start}=T_0+(r_0/c\beta)(1-\beta\cos(\max[0,\theta_v-\Delta\theta]))$
and ends at 
$T_{end}=T_0+(r_0/c\beta)(1-\beta\cos(\theta_v+\Delta\theta))$.
Here we set $T_0=0$.
({\it middle panel}): The pulse width $W_{FWHM}$ 
at half maximum in units of $r_0/c\beta\gamma^2$
is shown as a function of $(1+\gamma^2 \theta_v^2)$.
The pulse width $W_{FWHM}$ can be estimated as $W_{FWHM}\propto \delta \propto
(1+\gamma^2 \theta_v^2)$ for $\theta_v \sim \Delta \theta$
and $W_{FWHM}\propto \delta^{1/2} \propto (1+\gamma^2 \theta_v^2)^{1/2}
\sim \gamma \theta_v$ for $\theta_v \gg \Delta \theta$,
which are shown by dashed and dotted lines, respectively.
({\it lower panel}):
The product of the peak flux $\nu F_{\nu}^{peak}$ 
and the pulse width $W_{FWHM}$, in units of $2 A_0 r_0^2 \gamma \nu_0'/
\beta D^2$, is shown 
as a function of $(1+\gamma^2 \theta_v^2)$.
This quantity is about the total fluence $\nu S_{\nu}$ in Figure 1.
The peak flux $\nu F_{\nu}^{peak}$
can be estimated from the relation $\nu F_{\nu}^{peak} W_{FWHM}
\sim \nu S_{\nu} \propto \delta^{-1+\alpha_B} (\delta^{-1+\beta_B})$
when the maximum frequency $\nu_{max}$
is higher (lower) than the observed frequency,
which are shown by dashed (dotted) lines.
In all panels, we adopt $\gamma \Delta \theta=1$,
$\alpha_B=-1$, $\beta_B=-2.5$ and $s=1$, and 
the observed frequency is $\nu=200{\rm keV}(\gamma \nu_0'/10^3{\rm keV})$.
}
\end{figure}

\newpage 
\begin{figure}
    \epsfysize 17cm 
    \epsfbox{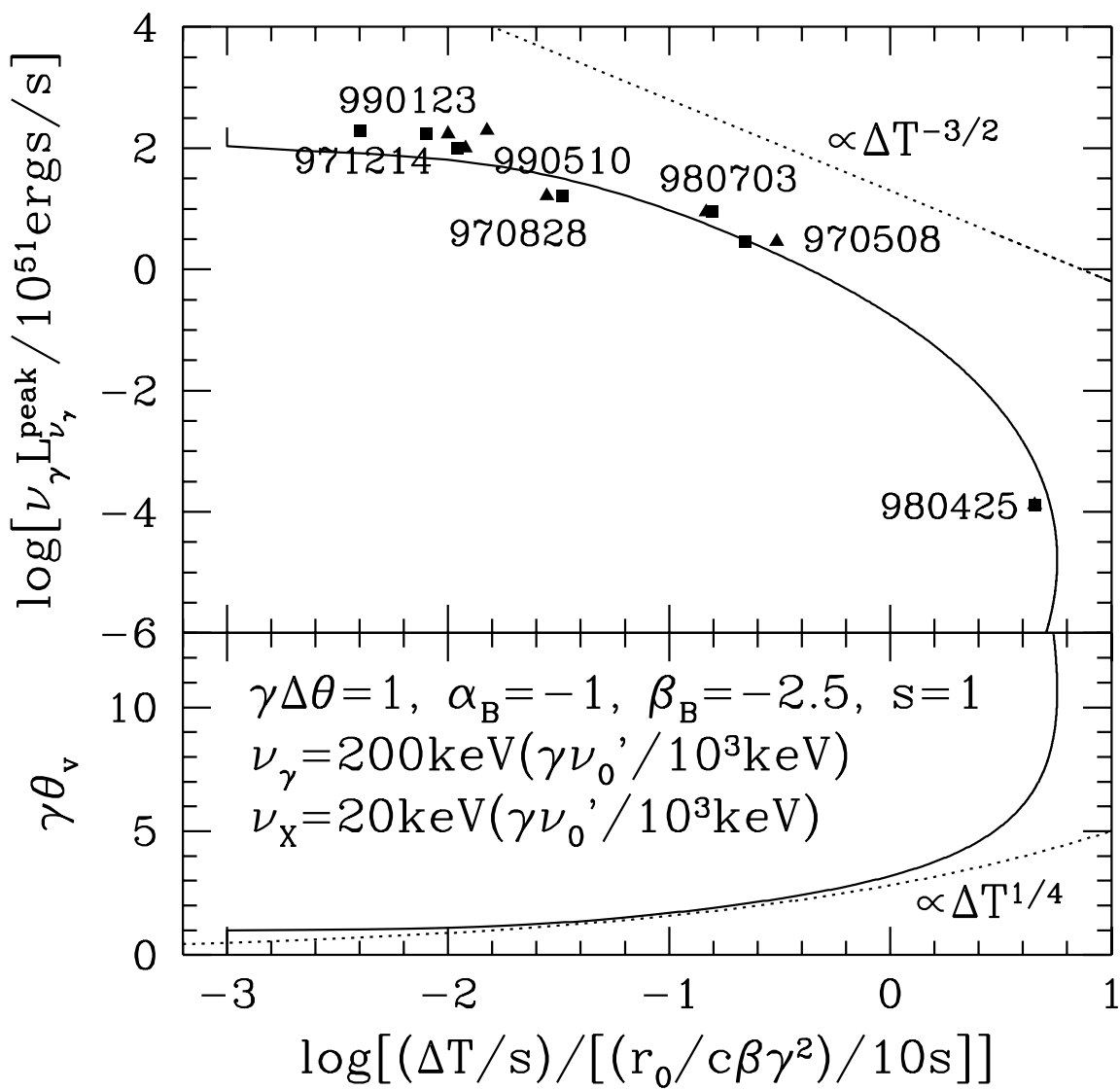}
\caption[fig3.ps]
%\figcaption
{The isotropic peak luminosity at frequency
$\nu_{\gamma}=200{\rm keV}(\gamma \nu_0'/10^3{\rm keV})$
is shown as a function of the spectral lag of the peak time $\Delta T$ 
by varying the viewing angle $\gamma \theta_v$.
The corresponding viewing angle $\gamma \theta_v$ is shown in the lower panel.
The spectral lag $\Delta T$ is defined by the difference of the peak time
between frequencies $\nu_{\gamma}=200{\rm keV}(\gamma \nu_0'/10^3{\rm keV})$
and $\nu_{X}=20{\rm keV}(\gamma \nu_0'/10^3{\rm keV})$.
We adopt $\gamma \Delta \theta=1$,
$\alpha_B=-1$, $\beta_B=-2.5$, $s=1$,
$r_0/c\beta \gamma^2=10$ s and an arbitrary vertical unit 
$2 c A_0 r_0 \gamma^2/D^2$.
All points with $\Delta T\le 10^{-4}(r_0/c \beta \gamma^2)$
are plotted at $\Delta T=10^{-4}(r_0/c \beta \gamma^2)$.
The observed luminosity-spectral lag relations for seven bursts
with known redshifts are shown from Table 1 
in Norris, Marani \& Bonnell (2000).
Squares and triangles are lags for regions down to $0.5$ and $0.1$
of the peak luminosity, respectively.
The dotted lines represent the power law behavior 
$\nu_{\gamma} L^{peak}_{\nu_{\gamma}} \propto 
\Delta T^{(-2+\alpha_B)/(s+1)}=
\Delta T^{-3/2}$
%from equation (\ref{eq:lagsubjet})
and $\gamma \theta_v \propto \Delta T^{1/2(s+1)}=
\Delta T^{1/4}$ (Ioka \& Nakamura 2001).
%from equation (\ref{eq:tp}).
}
\end{figure}

\newpage 
\begin{figure}
    \epsfysize 17cm 
    \epsfbox{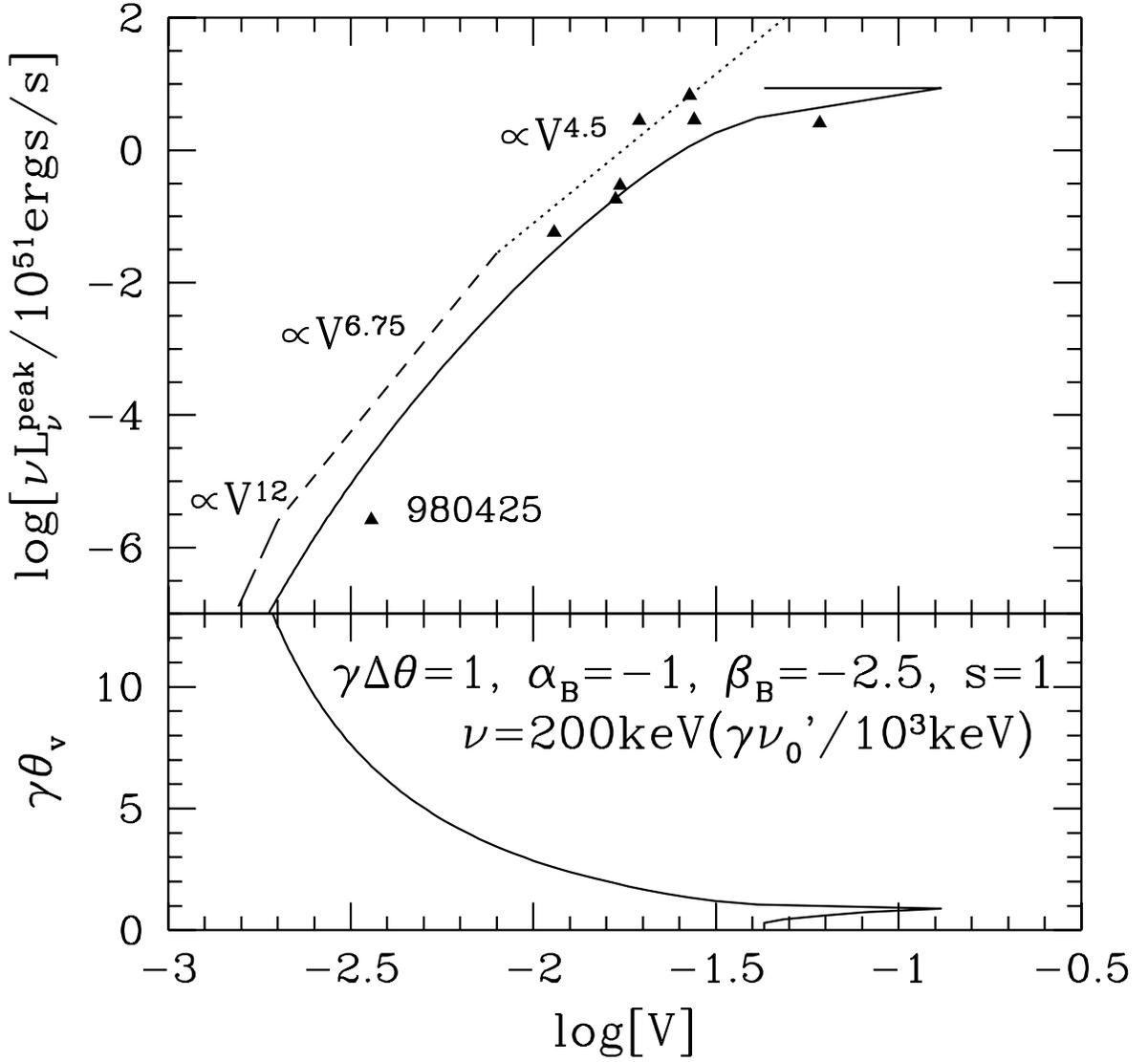}
\caption[fig4.ps]
%\figcaption
{The isotropic peak luminosity at frequency
$\nu=200{\rm keV}(\gamma \nu_0'/10^3{\rm keV})$
is shown as a function of the variability $V$ 
by varying the viewing angle $\gamma \theta_v$.
The corresponding viewing angle $\gamma \theta_v$ is shown in the lower panel.
We adopt $\gamma \Delta \theta=1$,
$\alpha_B=-1$, $\beta_B=-2.5$, $s=1$,
the power density spectrum of the GRB time histories $P\propto f^{d}
\sim f^{-5/3}$,
and appropriate off-sets.
The observed luminosity-variability relations for seven bursts
with known redshifts are shown by filled triangles  from Table 1 
of Fenimore \& Ramirez-Ruiz (2000).
The analytical estimates of
$\nu L^{peak}_{\nu} \propto V^{(-2+\alpha_B)/(d+1)}
\sim V^{4.5}$,
$\nu L^{peak}_{\nu} \propto V^{(-2+\beta_B)/(d+1)}
\sim V^{6.75}$, and
$\nu L^{peak}_{\nu} \propto V^{(-3+2\beta_B)/(d+1)}
\sim V^{12}$
are shown in each region
by dotted, dashed, and long dashed lines, respectively.
}
\end{figure}

%
% Tables
%

\end{document}